\documentclass{emulateapj}

\newcommand{\tacc}{\tau_{\rm a}}
\newcommand{\tmag}{\tau_{\rm m}}
\newcommand{\tint}{\tau_{\rm i}}

\newcommand{\mdotacc}{\dot M_{\rm a}}
\newcommand{\gammacrit}{\gamma_{\rm c}}

\newcommand{\rco}{R_{\rm co}}
\newcommand{\rt}{R_{\rm t}}

\newcommand{\rout}{R_{\rm out}}

\newcommand{\spineq}{T_{\rm eq}}
\newcommand{\omegaeq}{\Omega_*^{\rm eq}}

\shorttitle{Disk Locking, Revisited}
\shortauthors{Matt \& Pudritz}

\begin{document}


\title{Does Disk Locking Solve the Stellar Angular Momentum Problem?}

\author{Sean Matt\altaffilmark{1,2} and Ralph E. Pudritz\altaffilmark{1}}


\altaffiltext{1}{Physics \& Astronomy Department, McMaster University, 
Hamilton ON, Canada L8S 4M1\\
matt@physics.mcmaster.ca, pudritz@physics.mcmaster.ca}
\altaffiltext{2}{CITA National Fellow}

\begin{abstract}

We critically examine the theory of disk locking, which assumes that
the angular momentum deposited on an accreting protostar is exactly
removed by torques carried along magnetic field lines connecting the
star to the disk.  In this letter, we consider that the differential
rotation between the star and disk naturally leads to an opening
(i.e., disconnecting) of the magnetic field between the two.  We find
that this significantly reduces the spin-down torque on the star by
the disk.  Thus, disk-locking cannot account for the slow rotation
($\sim$10\% of breakup speed) observed in several systems and for
which the model was originally developed.


\end{abstract}

\keywords{accretion, accretion disks --- MHD --- stars: formation ---
stars: magnetic fields --- stars: pre-main-sequence --- stars:
rotation}

\section{Introduction \label{sec_intro}}



We now know that most protostars undergo a phase in which they accrete
mass from a disk and that a stellar magnetic field dominates this
process near the star.  However, the transport of angular momentum,
while fairly well-understood within the disk, remains a significant
challenge for models of stellar spin \citep[for a review,
see][]{bodenheimer95}.  Classical T Tauri stars (CTTS's) comprise a
class of accreting protostars for which there is an abundance of
observational data.  For typical parameters in these systems, the
angular momentum deposited by accretion can spin the star up to
breakup speed in $\sim 10^5$ years, assuming the star hadn't already
formed at breakup speed.  Since the accretion lifetime is often
greater than $10^6$ yr, the stars must somehow rid themselves of this
excess angular momentum.  Furthermore, a significant number of CTTS's
(the so called ``slow rotators'') have spin rates as slow as
$\sim$10\% of breakup speed \citep[e.g.,][]{bouvierea93}, requiring
significant torques to maintain.

\citet[][hereafter K91]{konigl91} proposed the currently accepted view
that the slow rotators could be explained with a ``disk-locking'' (DL)
mechanism \citep[adapted from][]{ghoshlamb79}, for which the star
needs a kilogauss-strength dipole magnetic field.  Indeed, CTTS's are
now known to possess kG-strength fields
\citep*[e.g.,][]{johnskrull3ea99}, and magnetic star-disk interaction
models have also explained several observed spectral features of
CTTS's (e.g., K91).  However, observations by \citet{stassunea99} and
\citet{johnskrullea99} and theoretical considerations of
\citet{safier98} have called the standard DL scenario into question.
According to DL theory, magnetic field lines causally connect the star
to the disk (acting as ``lever arms'') and carry torques that oppose
and balance the angular momentum deposited by accretion.  Although not
usually considered in DL theory, \citet*[][hereafter
UKL]{uzdensky3ea02} demonstrated that differential rotation between
the star and disk leads to an opening of the field, drastically
reducing the magnetic connection between the two.  In this letter, we
consider the consequences of this reduced connection on DL.  We find
that the magnetic torque is significantly reduced, and show that the
DL model therefore cannot account for the angular momentum loss of the
slow rotators.


\section{Torques in the Star-Disk Interaction} \label{sec_torques}

\begin{figure}
\epsscale{1.1}
\plotone{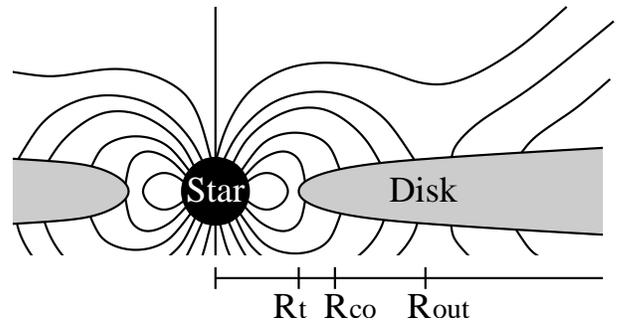}

\caption{Magnetic star-disk interaction.
\label{fig_cartoon}}

\end{figure}

In order to discuss the disk-locked state, in which magnetic torques
on the star exactly balance the angular momentum deposited by
accreting material, we must first discuss the general case for an
arbitrary state of the star-disk system but with the proper inclusion
of field line opening.  To this end, we mainly follow the formulation
of \citet[][hereafter AC96]{armitageclarke96}, and a more detailed
presentation of our model will be given in \citet{mattpudritz04}.
According to the typical assumptions, the central star contains a
rotation-axis-aligned dipolar magnetic field.  This field is anchored
in the stellar surface and also threads a thin, Keplerian accretion
disk, in which the kinetic energy of orbiting gas is much greater than
the magnetic field energy.  The region above the star and disk
contains low-density gas and is dominated by the magnetic field.  The
system is assumed to exist in a steady-state, where the accretion rate
$\mdotacc$ is constant in time and at all radii in the disk.  Figure
\ref{fig_cartoon} illustrates the basic idea and identifies the
location where the disk is truncated ($\rt$) and the corotation radius
($\rco$), where the Keplerian angular rotation rate equals that of the
star.  The region in the disk threaded by closed stellar field lines
is delimited by $\rout$ (the usual assumption is $\rout \gg \rco$).

The assumption of a steady-state $\mdotacc$ requires that the net
angular momentum carried away from each annulus of the disk (of width
$\delta r$ and vertically integrated) equals
\begin{eqnarray}
\label{eqn_tacc}
\delta \tacc = 0.5 \; \mdotacc \sqrt{G M_*} \; r^{-1/2} \; \delta r,
\end{eqnarray}
where $M_*$ is the stellar mass.  If the torque anywhere differs from
this, the disk will restructure itself on an dynamical timescale,
reestablishing the steady-state.

The star and disk rotate at different angular speeds, except at the
singular location $\rco$.  Thus the magnetic field becomes twisted
azimuthally by differential rotation, and magnetic forces act to
restore the dipole configuration---conveying torques between the star
and disk.  The magnetic torque on the star exerted by a bundle of
field lines threading an annulus in the disk, can be written as (AC96)
\begin{eqnarray}
\label{eqn_dtmag}
\delta \tmag = \gamma \mu^2 r^{-4} \delta r,
~~~~~~ ~~~~ ~~~~~~
\gamma \equiv B_\phi / B_z.
\end{eqnarray}
This differential torque depends only on the strength of the dipole
moment $\mu$ and on $\gamma$, which is the ``twist'' (or pitch angle)
of the field.  The sign of the torque, here and in all equations, is
relative to the star, so a positive torque spins the star up and thus
spins the disk down.

The magnetic field cannot be perfectly frozen-in to the gas of the
disk.  Instead, it diffuses or reconnects through the disk azimuthally
at some speed, $v_{\rm d} \approx \gamma \; \eta_{\rm t}/h$ (UKL),
where $\eta_{\rm t}$ is the magnetic diffusivity (we use the subscript
``t'' to suggest a turbulent value) and $h$ is the local, vertical
scale height of the disk.  For a standard $\alpha$-disk
\citep{shakurasunyaev73}, $v_{\rm d} = \beta v_{\rm kep} \gamma$,
where $v_{\rm kep}$ is the orbital speed, $\beta \equiv (\alpha /
P_{\rm t}) (h/r)$, and $P_{\rm t}$ is the turbulent Prandtl number (=
turbulent viscosity divided by $\eta_{\rm t}$).  Since both $h/r$ and
$\alpha / P_{\rm t}$ are likely to have weak (though unknown)
dependences on $r$, we assume $\beta$ is constant.  The radial
dependence of the magnetic torque is mainly dominated by the quick
falloff of the dipole magnetic field ($r^{-3}$), and so a small radial
dependence of $\beta$ will not much affect our results (AC96).

In general, $\beta$ is a simple scale factor that compares $v_{\rm d}$
to $v_{\rm kep}$.  It measures the coupling of the magnetic field to
the disk: $\beta \ll 1$ corresponds to strong coupling and $\beta \gg
1$ to weak.  The value of $\beta$ is unknown (AC96 used $\beta = 1$).
Typical $\alpha$-disk parameters give a limit of $\beta \le 1$ and a
likely value of a few orders of magnitude lower.  For reasonable
fiducial parameters, $\beta = 10^{-2} (\eta_{\rm t} / 10^{16}$ cm$^2$
s$^{-1}) (R_\odot / h) (100$ km s$^{-1} / v_{\rm kep})$.  However,
given the uncertainties, we keep $\beta$ as a free parameter.

Where the field connects the star and disk, the magnetic twist (and so
$v_{\rm d}$) will increase, until the field can slip through the disk
at a rate equal to the differential rotation rate.  Thus, we expect a
steady-state in which
\begin{eqnarray}
\label{eqn_gamma}
\gamma = \beta^{-1} \left[{(r / \rco)^{3/2} - 1}\right].
\end{eqnarray}
Note that $\gamma(\rco) = 0$, since the differential rotation is zero
there.  For $r > \rco$, the twist increases to infinity.

Several recent theoretical and numerical studies of the star-disk
interaction \citep[see UKL and references
therein;][]{lyndenbellboily94, agapitoupapaloizou00} have shown with
certainty that, as a dipole field is twisted this way, the magnetic
torque reaches a maximum for finite $\gamma$, but then reduces back to
zero for larger values.  This occurs because, when the magnetic field
is twisted enough so that the magnetic pressure associated with
$B_\phi$ overcomes the poloidal field lines, the latter will
``inflate,'' opening to infinity at mid latitudes.  The star and disk
then become causally disconnected because open field lines cannot
convey torques between the two.  The critical twist at which this
happens is very nearly $\gamma = \gammacrit \approx 1$ (e.g., UKL).

The opening of field lines by this process can be added to the theory.
For simplicity, we assume that the star is connected to the disk, as
described above, at all locations where the twist is less than the
critical value $\gammacrit$.  Where $\gamma \ge \gammacrit$, the disk
and star are disconnected, and so the differential magnetic torque
there is zero.  In other words, this process determines the outer
radius of closed stellar field lines, $\rout$.  From equation
\ref{eqn_gamma}, we find that $\rout = (1 + \beta \gammacrit)^{2/3}
\rco$.  AC96 assumed that the star is connected to the disk over a
very large radial extent ($\rout \rightarrow \infty$), which is
equivalent to the assumption that $\gammacrit \rightarrow \infty$.

\begin{figure}
\epsscale{1.1}
\plotone{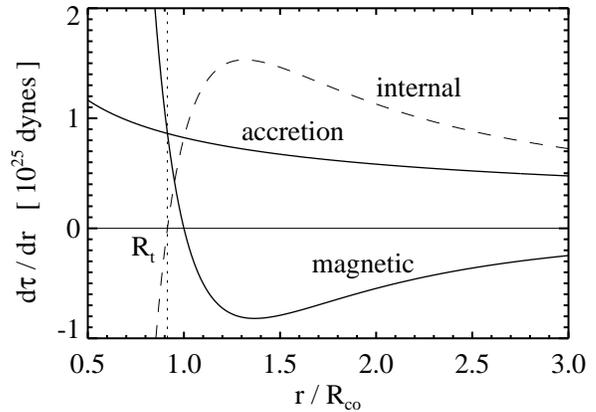}

\caption{Differential torques in the disk midplane predicted for
$\beta = 1$ and $\gammacrit \rightarrow \infty$, for BP Tau (see
text).  The system is shown in the equilibrium state, where the net
torque on the star is zero, requiring a stellar rotation period of 7.5
d (so $\rco \approx 6.4 R_*$).
\label{fig_dtorques}}

\end{figure}

A combination of equations \ref{eqn_dtmag} and \ref{eqn_gamma} reveals
the radial dependence of the differential magnetic torque ($\delta
\tmag$).  Figure \ref{fig_dtorques} shows all differential torques
(per $\delta r$) in the disk, as a function of radius (normalized to
$\rco$), assuming $\beta = 1$ and $\gammacrit \rightarrow \infty$
(i.e., the AC96 solution).  For the Figure, we adopt representative
system parameters from the well-studied CTTS BP Tau: $\mdotacc = 3
\times 10^{-8} M_\odot$ yr$^{-1}$, $R_* = 2 R_\odot$, and $M_* = 0.5
M_\odot$ \citep{gullbringea98}, and $B_*$ = 2 kG
\citep{johnskrull3ea99}.  The stellar rotation period is 7.5 days (see
\S \ref{sec_equilib}).

The lines in Figure \ref{fig_dtorques} labelled ``accretion'' and
``magnetic'' represent the differential accretion and magnetic
torques, $\delta \tacc$ and $\delta \tmag$, from equations
\ref{eqn_tacc} and \ref{eqn_dtmag}, respectively.  The steady-state
condition requires the disk to structure itself such that the net
differential torque everywhere equals $\delta \tacc$.  So when
external magnetic torques ($\delta \tmag$) act on the disk, there are
torques internal to the disk \citep[e.g., via the magnetorotational
instability,][]{balbushawley91} to counteract them and always provide
a differential torque defined by $\delta \tint \equiv \delta \tacc -
\delta \tmag$, represented by the dashed line in Figure
\ref{fig_dtorques}.  From the figure, it is evident that $\delta
\tmag$ is strongest near the star, where the magnetic field is strong,
and $\delta \tmag$ acts to spin up the star inside $\rco$.  At $\rco$,
$\delta \tmag$ goes to zero, since the field is not twisted there
(eq.\ \ref{eqn_gamma}).  Beyond $\rco$, the twist increases, and so
$\delta \tmag$ becomes stronger, now acting to spin down the star.
Since the dipole field strength decreases faster than the increase of
$\gamma$, $\delta \tmag$ reaches a maximum (in absolute value) and
then approaches zero as $r \rightarrow \infty$.

At the location where $\delta \tmag = \delta \tacc$ (and $\delta \tint
= 0$), the torque from the stellar magnetic field is all that is
neccesary to provide $\mdotacc$.  Thus, all of the specific angular
momentum of the disk material at that location will end up on the
star.  This defines the radius $\rt$ (dotted line in Fig.\
\ref{fig_dtorques}), where the disk is truncated, and from where
accretion will be magnetically channelled along field lines onto the
star (K91; not shown in Fig.\ \ref{fig_cartoon}).  So the net torque
on the star from the accretion of disk material, $\tacc$, is obtained
by integrating equation \ref{eqn_tacc} from $\rt$ to the surface of
the star.  Similarly, the net magnetic torque, $\tmag$, is obtained by
integrating equation \ref{eqn_dtmag} over the entire magnetically
connected region of the disk, from $\rt$ to $\rout$, which gives
\begin{eqnarray}
\label{eqn_tmag}
\tmag =  \mu^2 (3 \beta)^{-1} \rco^{-3}
  & \left[{2 \left({1 + \beta \gammacrit} \right)^{-1} 
  -  \left({1 + \beta \gammacrit}\right)^{-2}} \right. \nonumber \\
  & \left.{-2 (\rco / \rt)^{3/2}  
  +   (\rco / \rt)^{3}} \right].
\end{eqnarray}
This is exactly the solution found by AC96 for the special case that
$\beta = 1$ and $\gammacrit \rightarrow \infty$.  Our formulation
allows for an arbitrary value of the diffusion parameter $\beta$ and
includes the effects of field line opening via twisting to a critical
value of $\gammacrit$.

We determine the dependence of the spin-down torque on $\beta$ by
setting $\rt = \rco$ in equation \ref{eqn_tmag} and adopting
$\gammacrit = 1$ (as justified by, e.g., UKL).  We find that $\tmag
\propto \beta$ in the strong coupling limit ($\beta \ll 1$), has a
maximal value for $\beta = 1$, and $\tmag \propto \beta^{-1}$ in the
case of weak coupling ($\beta \gg 1$).  This behavior can be
understood as a competition between two different effects: one is
that, when $\beta$ is small, the size of the magnetically connected
region in the disk ($\rout$) is small, reducing $\tmag$; secondly,
when $\beta$ is large, the field is less twisted (eq.\
\ref{eqn_gamma}) at each radius, reducing $\delta \tmag$.  These two
effects conspire to give a maximal value of $\tmag$, for the critical
value of $\beta = 1$, which thus represents the ``best case'' for DL
theory.  Even for this ``best case,'' the proper treatment of field
line opening ($\gammacrit = 1$) gives a total spin-down torque that is
a factor of four times less than if $\gammacrit \rightarrow \infty$.

\begin{figure}
\epsscale{1.1}
\plotone{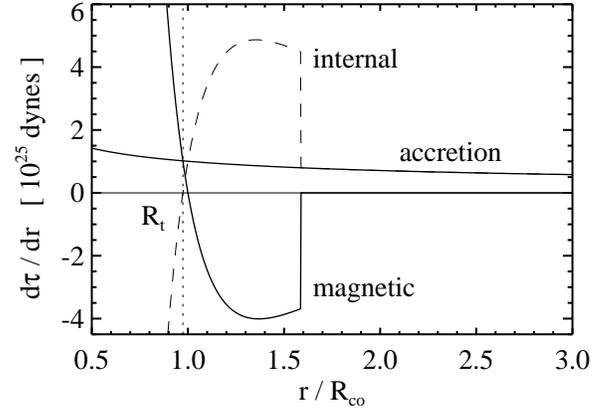}

\caption{Same as Figure \ref{fig_dtorques}, except that $\gammacrit =
1$, so the star is magnetically disconnected from the disk beyond
$\rout \approx 1.6 \rco$, and the equilibrium rotation period is 4.1 d
($\rco \approx 4.3 R_*$).
\label{fig_dtorques2}}

\end{figure}

To illustrate the effect of field line opening, Figure
\ref{fig_dtorques2} shows the differential torques in the disk, for
the same parameters as Figure \ref{fig_dtorques}, except that now
$\gammacrit = 1$, and the spin period is 4.1 d (see \S
\ref{sec_equilib}).  The magnetic field is now open for $r \ge \rout
\approx 1.6 \rco$, and $\delta \tmag / \delta r$ is zero there (by
assumption).  It is clear that the integrated torque $\tmag$ will be
less than if the magnetic field were everywhere closed (i.e., for
$\gammacrit \rightarrow \infty$).

\section{The Disk-Locked State} \label{sec_equilib}

For any given values of $M_*$, $R_*$, $B_*$, $\mdotacc$, and the
stellar rotation period, the star-disk interaction theory presented in
section \ref{sec_torques} gives the net torque ($\tacc + \tmag$) on
the star.  The system is stable in that, for fast rotation, the net
torque spins the star down, and for slow rotation, the star spins up.
For typical CTTS parameters, the net torque will spin the star up or
down in $\sim 10^5$ yr (e.g., AC96). So one expects that most systems
older than this will exist in an equilibrium state, in which the net
torque on the star is zero, and in which the system is ``disk locked''
with a period $\spineq$.

The equilibrium spin state is determined by combining the condition of
net zero torque ($\tacc + \tmag = 0$) with the definition of $\rt$
(i.e., where $\delta \tacc = \delta \tmag$).  One finds that, in the
DL state, there is a single predicted value of the truncation radius,
$\rt^{\rm eq}$, that depends on the quantity $\beta \gammacrit$, but
not on any other system parameters.  If $\beta \gammacrit$ approaches
zero, $\rt^{\rm eq}$ approaches $\rco$.  The ratio $\rt^{\rm eq}/\rco$
monotonically decreases for increasing $\beta \gammacrit$, but it
never becomes smaller than 0.91, as $\beta \gammacrit
\rightarrow \infty$.

We calculate that the equilibrium angular spin rate of the star is
given by
\begin{eqnarray}
\label{eqn_spineq}
\omegaeq = C(\beta, \gammacrit) \;
  \mdotacc^{3/ 7} \left({G M_*}\right)^{5/ 7} 
  B_*^{-{6/ 7}} R_*^{-{18/ 7}},
\end{eqnarray}
where $C(\beta, \gammacrit)$ is a dimensionless function that depends
only $\beta$ and $\gammacrit$.

Figures \ref{fig_dtorques} and \ref{fig_dtorques2} show the torques
for stars that are in their equilibrium spin states, $\spineq = 7.5$
and 4.1 d, respectively.  A comparison of these two Figures reveals
the effect of field line opening (see \S \ref{sec_torques}).  When
this occurs, the net spin-down torque is decreased, so the star must
spin faster for it to be in equilibrium.  A faster spin reduces
$\rco$, so the torques come from closer to the star, where the dipole
field is stronger.  Thus, a faster spin makes up for the decreased
size of the magnetically connected region.  Also, the truncation
radius is nearer the corotation radius in Figure \ref{fig_dtorques2}
than in \ref{fig_dtorques} ($\rt^{\rm eq} \approx 0.97 \rco$, compared
to $\approx 0.91 \rco$).

\begin{figure}
\epsscale{1.1}
\plotone{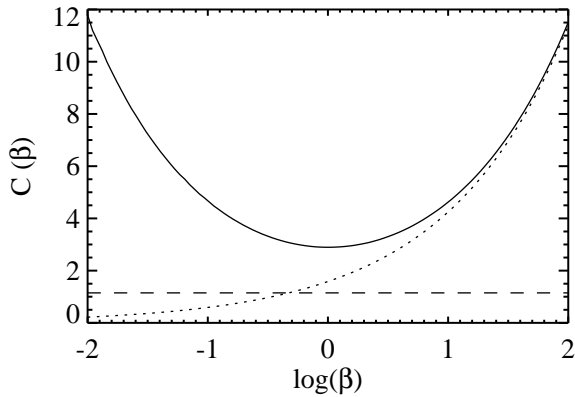}

\caption{Spin-rate factor $C$ of equation \ref{eqn_spineq}, as a
function of $\log(\beta)$.  The solid line represents the model with
field line opening ($\gammacrit = 1$).  For comparison, the dotted
line represents $\gammacrit \rightarrow \infty$, while the dashed line
is from K91.
\label{fig_cbg}}

\end{figure}

Equation \ref{eqn_spineq} has the same dependences on system
parameters as several DL models in the literature \citep[e.g., K91
and][]{ostrikershu95}, but the factor $C$ varies slightly from model
to model.  For example, both K91 and \citet{ostrikershu95} use $C
\approx 1.1$, while the AC96 value (i.e., for $\beta = 1$ and
$\gammacrit \rightarrow \infty$) is $C \approx 1.6$.  Our formulation
of the problem includes the effects of field line opening and
determines the dependence of the function $C$ on the diffusion
parameter $\beta$.  The solid line of Figure \ref{fig_cbg} reveals
this dependence, for $\gammacrit = 1$.  For comparison, if the field
remains closed for arbitrary twist values ($\gammacrit \rightarrow
\infty$), $C(\beta) \approx 1.6 \beta^{3/7}$ (dotted line). The K91
value of $C$ is indicated by a dashed line.  The effect of field line
opening is significant.  For the ``best case'' ($\beta = 1$), the
predicted spin rate is 1.8 times faster than for $\gammacrit
\rightarrow \infty$ and 2.6 times faster than predicted by K91.  For a
more reasonable value of $\beta = 10^{-2}$, the predicted spin is more
than an order of magnitude faster than all other DL models, in which
the field is typically assumed to remain closed while largely twisted.

To date, DL models have had some success in explaining the spin rates
of slow rotators.  For the example case of BP Tau (see \S
\ref{sec_torques}), when one assumes $\beta = 1$ and $\gammacrit
\rightarrow \infty$, equation \ref{eqn_spineq} predicts $\spineq =
7.5$ d, corresponding to 6\% of breakup speed and agreeing with the
predictions of most models in the literature.  This predicted period
is remarkably similar to the observed value of 7.6 d \citep{vrbaea86}.
However, when one properly includes the effect of field line opening
(for $\gammacrit = 1$), the ``best case'' ($\beta = 1$) model predicts
$\spineq = 4.1$ d, and the spin-up time is less than $10^5$
yr---significantly shorter than BP Tau's age of $6 \times 10^5$ yr
\citep{gullbringea98}.  Further, for $\beta = 0.01$, $\spineq = 1.0$
d, with a spin-up time of $5 \times 10^5$ yr!


It is also important to note that we have thus far used parameters for
BP Tau determined by \citet{gullbringea98} and
\citet{johnskrull3ea99}.  If instead we use the higher $\mdotacc$
derived by \citet{hartigan3ea95} of $1.6 \times 10^{-7} M_\odot$
yr$^{-1}$, even the ``standard'' ($\beta = 1$, $\gammacrit \rightarrow
\infty$) model predicts $\spineq = 3.6$ d, with a spin-up time of $2
\times 10^4$ yr.  Alternatively, if we use the $3 \sigma$ upper limit
on the dipole field strength of 200 G from \citet{johnskrullea99},
even the ``standard'' theory gives $\spineq = 1.0$ d.

We must conclude that either BP Tau and other slow rotators are
rapidly spinning up, or the DL picture, as an explanation of stellar
spin, is incomplete.  The former requires (e.g.)\ a recent, dramatic
increase in $\mdotacc$---this seems unlikely, especially for BP Tau,
for which the accretion disk is probably in the process of clearing
out \citep{dutrey3ea03}.  In order for BP Tau to currently be in an
equilibrium spin state, there must be significant spin-down torques on
the star other than those carried along field lines connecting the
star to the disk.


\section{Summary of Problems With DL} \label{sec_summary}


We have shown that a large portion of the magnetic field connecting
the star to the disk will open up due to the differential rotation
between the two, resulting in a spin-down torque on the star by the
disk that is much less than if the field is assumed to remain closed.
The predicted disk-locked spin rate is therefore much faster, so that
the DL scenario cannot explain the angular momentum loss of the slow
rotators.


In addition, there are at least three, completely independent issues
raised by other authors: 1) \citet{stassunea99} found no correlation
between accretion parameters and spin rates of TTS in Orion.  2)
CTTS's apparently do not have strong dipole fields
\citep[e.g.,][]{safier98, johnskrullea99}, which are required for DL
at slow spin rates.  3) Stellar winds are expected to open field lines
that would otherwise connect to the disk \citep{safier98}.  A disk
wind could have a similar effect.






We conclude that, in order for accreting protostars to spin as slowly
as 10\% of breakup speed, there must be spin-down torques acting on
the star other than those carried by magnetic field lines connecting
the star to the disk.  The presence of open stellar field lines
implies that excess angular momentum may be carried by a stellar wind
\citep[e.g.,][]{toutpringle92}, but this remains an open question.  We
are investigating the role of stellar winds, since then $\spineq$ will
depend mostly upon the stellar wind parameters, thereby explaining the
lack of correlation between spin periods and accretion parameters.

\acknowledgements

We are grateful for discussions in sunny weather with Cathy Clarke and
Bob Mathieu. This research was supported by NSERC of Canada and CITA.





\end{document}